\documentclass[epj]{svjour}
\usepackage{graphicx}
\usepackage{dcolumn}
\usepackage{bm}
\usepackage{hyperref}
\usepackage{amsmath}
\begin{document}
\title{Roles of cooperative effects and disorder in photon localization: The case of a vector radiation field}

\author{L. Bellando\inst{1}\inst{2} \and A. Gero\inst{3}\inst{4}\and E. Akkermans\inst{3}\and R. Kaiser\inst{2} % etc
% \thanks is optional - remove next line if not needed
%\thanks{\emph{Present address:} Insert the address here if needed}%
}                     % Do not remove
%
%\offprints{}          % Insert a name or remove this line
%
\institute{Universit\'{e} de Bordeaux, CNRS, LOMA, UMR 5798, F-33405 Talence, France \and Universit\'{e} C\^{o}te d'Azur, CNRS, Institut de Physique de Nice, F-06560 Valbonne, France \and Department of Physics, Technion - Israel Institute of Technology, 32000 Haifa, Israel \and Department of Education in Technology and Science, Technion - Israel Institute of Technology, 32000 Haifa, Israel }
\date{}
% The correct dates will be entered by Springer
%
\abstract{We numerically study photon escape rates from  three-dimensional atomic gases and investigate the respective roles of cooperative effects and disorder in photon localization, while taking  into account the vectorial nature of light. A scaling behavior is observed for the escape rates, and photons undergo a crossover from delocalization toward  localization as the optical thickness of the cloud is increased. This result indicates that light localization is dominated by cooperative effects rather than disorder. We compare our results with those obtained in the case of a scalar radiation field and find no significant differences. We conclude that the scalar model constitutes an excellent approximation when considering  photon escape rates from atomic gases.
 % end of PACS codes
} %end of abstract
\maketitle
\section{Introduction}

Light localization in cold atomic gases, namely an overall decrease of photon escape rates from the cloud, is a subject of interest in atomic and optical physics, both theoretically and experimentally \cite{skip1,BGAK,skip2,skip4,Guerin2016}.
Localization of light in atomic media can be induced by two mechanisms. The first one, Anderson localization, originates from interference effects in the presence of disorder \cite{anderson}, whereas the second mechanism,  cooperativity, stems from the synchronization of atomic dipoles and does not require disorder \cite{dicke}. To investigate photon localization, resulting from the contribution of these two mechanisms, two different approaches have been proposed.

The first approach studies the complex spectrum of the effective Hamiltonian describing the atomic system. There, the real part of an eigenvalue corresponds to the
energy of the eigenstate, while its imaginary part is related to the decay rate \cite{BGAK,orlowski,svidzinsky2,bienaime}. A numerical analysis based on this approach has shown that localization of light can be achieved in random three-dimensional atomic media only for a scalar radiation field; it cannot be achieved when the vectorial properties of the electromagnetic wave are taken into account  \cite{skip1,BGAK,skip2}. A similar result has been obtained in two-dimensional systems as well \cite{Loc2DBachelar}. Current studies following this approach examine the influence of external laser \cite{Guerin2016,bienaime,Araujo2016,Roof2016,SubVSRT} and magnetic \cite{SkipetrovPRA} fields or additional disorder  \cite{Celardo2017} on photon emission rates.

According to the second approach, photon escape rates are determined by the time evolution of the ground-state population obtained from the reduced atomic density matrix of the gas. This time evolution is governed by the spectrum of the imaginary part of the effective Hamiltonian \cite{ernst,tallet1,tallet2,PRL2}. Unlike the first approach, the second one takes into account for detection purposes only the on-shell photons that leave the cloud. Studies based on the latter approach have been restricted to the interaction of  atoms with a scalar radiation field. These  studies have compared the roles of disorder and cooperative effects, e.g., superradiance and subradiance, in photon localization. It has been shown that in  two- and three-dimensional media, photon localization occurs as a crossover between two limits: the single-atom limit where photons are spatially delocalized  and spontaneous emission of independent atoms occurs, and the opposite limit where the photons are trapped in the gas for a very long time \cite{PRL2,PRA2}. As these two limits are connected by a crossover rather than a disorder-driven phase transition as expected from Anderson localization \cite{gang}, one can argue that photon localization is dominated by cooperative effects rather than disorder. In one-dimensional atomic gases, the single-atom limit is never reached and photons are always localized in the cloud - a result clearly attributed to cooperative effects \cite{EPL}.

The study described here examines whether the significant difference between the scalar and vector models in regard to photon localization, obtained within the framework of the first approach and mentioned above, occurs as part of the second approach as well. In other words, we focus on the emission of (detectable) photons from three-dimensional atomic gases in the case of a vector radiation field and compare the findings to those of the scalar model. We show that the vector and scalar models qualitatively lead to  the same results. Moreover, in both cases a scaling behavior is observed for the escape rates, and  photons undergo a crossover from delocalization toward  localization as the optical thickness of the cloud is increased. The last result suggests that photon localization is dominated by cooperative effects rather than disorder. This conclusion is further validated by the finding that a similar scaling behavior is obtained for ordered media. Concerning  photon localization, we think that the similarity between the scalar and vector models  has an important impact on both theory and experiments.

The paper is organized as follows. In Section \ref{Mod}, we present the model  of a gas of identical atoms interacting with the radiation field. In Section \ref{Heff}, the atomic effective Hamiltonian is introduced, and in Section \ref{CER} the photon collective emission rates from the atomic cloud are derived. In Section \ref {Met} we describe our numerical methods, and  in Sections \ref{Res1}-\ref{Res2} present our findings. We finally discuss our results in Section \ref{discuss} and draw some conclusions in Section \ref{conclusions}.

\section{Model}
\label{Mod}

Atoms are taken as identical two-level systems, a ground state $|g\rangle=|J_g=0, m_g=0\rangle$ and an excited state $|e\rangle=|J_e=1, m_e=0,\pm1\rangle$. $J$ is the quantum number of the total angular momentum and $m$ is its  projection
on a quantization axis, taken as the $\hat{z}$ axis. The states associated to $m_e$ are spanned by the Cartesian basis $\{|\alpha\rangle\}=\{|x\rangle, |y\rangle, |z\rangle\}$, where $|x\rangle=\frac{1}{\sqrt{2}}(|-1\rangle-|+1\rangle)$, $|y\rangle=\frac{i}{\sqrt{2}}(|-1\rangle+|+1\rangle)$ and $|z\rangle=|0\rangle$. Therefore, further on we denote the excited state by $|e_\alpha\rangle$.  The energy separation between the ground and excited states, including radiative shift, is $\hbar\omega_0$ and the natural width of the excited state is $\hbar\Gamma_0$.

In order to describe the dynamics of $N \gg 1$ atoms, distributed at random positions $\textbf{r}_i$ in an external radiation field, we use the Hamiltonian
$H=H_0+V$.
$H_0$ is the free Hamiltonian, written as a sum of the atomic and radiation terms,
\begin{equation}
H_0=\hbar\omega_{0}\sum_{i=1}^{N}\sum_{\alpha}|e_{\alpha}^{i}\rangle\langle e_{\alpha}^{i}|+\sum_{\textbf{k}\hat{\varepsilon}}\hbar\omega_{k}a^{\dag}_{\textbf{k}\hat{\varepsilon}}a_{\textbf{k}\hat{\varepsilon}}.
\label{H0}
\end{equation}
 Here, $a^{\dag}_{\textbf{k}\hat{\varepsilon}}$ ($a_{\textbf{k}\hat{\varepsilon}}$) is the creation (annihilation) operator of a photon of wave vector $\text{\textbf{k}}$ ($\omega_k=c|\text{\textbf{k}}|$)  and polarization $\hat{\varepsilon}$ ($\textbf{k}\cdot\hat{\varepsilon}=0$).

The light-matter interaction $V$ expressed in the electric dipole approximation is
\begin{equation}
V=-\sum_{i=1}^{N}\textbf{d}_{i}\cdot\textbf{E}(\textbf{r}_{i}),
\end{equation}
where  $\textbf{d}_{i}$ is the electric dipole moment operator of the
$i$-th atom.

The atoms are coupled to each other via the electric field operator:
\begin{equation}
\textbf{E(r)}=i\sum_{\textbf{k}\hat{\varepsilon}}\sqrt{\frac{\hbar\omega_{k}}{2\epsilon_0\mathcal{V}}}\left(a_{\textbf{k}\hat{\varepsilon}}\hat{\varepsilon}e^{i\textbf{k}\cdot\textbf{r}}-a^{\dag}_{\textbf{k}\hat{\varepsilon}}\hat{\varepsilon}^{*}e^{-i\textbf{k}\cdot\textbf{r}}\right).
\end{equation}

It is assumed that the typical speed of the atoms  is small compared to
$\Gamma_{0}/k_0$ but large compared to $\hbar k_0/ \mu$, where $\mu$ is
the mass of the atom and $k_0=\omega_0/c$ is the wavenumber of the light used to probe the system. Under this assumption, the
Doppler shift and recoil effects are negligible. It should be noted that recent theoretical and experimental studies have shown that under the  assumption above, subradiant states are robust against thermal decoherence  \cite{bienaime,Weiss}. Furthermore, we assume that the typical speed of the atoms is well above the recoil speed, so that their de-Broglie wavelength is much smaller than $2\pi/k_0$. Therefore, quantum statistical effects (e.g., Bose-Einstein condensation) do not play a role. Finally, we neglect retardation effects, so that interaction between atoms is instantaneous.

\section{Effective Hamiltonian}
\label{Heff}

Performing the trace over the radiation degrees of freedom of $H$, in the case of a single excitation, produces the non-Hermitian effective Hamiltonian \cite{PRL2,zoller}:
\begin{equation}
\begin{split}
H_{eff}= \left( \hbar\omega_0-i \frac{\hbar\Gamma_{0}}{2}\right) \sum_{i=1}^{N}\sum_{\alpha}|e_{\alpha}^{i}\rangle\langle e_{\alpha}^{i}|\\
-\frac{\hbar\Gamma_0}{2}\sum_{i\neq j}\sum_{\alpha\beta}g_{\alpha\beta}(\textbf{r}_{ij})S_{i,\alpha}^{(+)}S_{j,\beta}^{(-)},
\end{split}
\label{equ:effectiveHamiltonian}
\end{equation}
where $S_{i,\alpha}^{(+)}=|e_{\alpha}^{i}\rangle\langle g^i|$ is the raising operator of the $i$-th atom along the $\alpha$ direction and $S_{i,\alpha}^{(-)}=[S_{i,\alpha}^{(+)}]^{\dag}$ is the corresponding lowering operator.
The off-diagonal terms, $g_{\alpha\beta}(\textbf{r}_{ij})$, represent the resonant dipole-dipole interaction \cite{stephen,lehmberg,milonni} also obtained in classical electrodynamics \cite{jackson}:
\begin{equation}
\begin{split}
g_{\alpha\beta}(\textbf{r}_{ij})=\frac{3}{2} e^{ik_0r_{ij}}\bigg[\left(\frac{1}{k_0r_{ij}}+\frac{i}{(k_0r_{ij})^2}-\frac{1}{(k_0r_{ij})^3}\right)\delta_{\alpha\beta}\\-\left(\frac{1}{k_0r_{ij}}+\frac{3i}{(k_0r_{ij})^2}-\frac{3}{(k_0r_{ij})^3}\right)\frac{r_{ij,\alpha}r_{ij,\beta}}{r_{ij}^{2}}\bigg],
\end{split}
\label{equ:couplingvect}
\end{equation}
where $r_{ij}=|\textbf{r}_{i}-\textbf{r}_{j}|$  and $r_{ij,\alpha}$ is the projection of  $\textbf{r}_{ij}$ on the $\alpha$ direction.
The imaginary part of the resonant dipole-dipole interaction describes contribution from energy conserving processes, while its real part originates from virtual processes  \cite{svidzinsky2}.

Averaging (\ref{equ:couplingvect}) over the random orientations of the pairs of atoms gives
\begin{equation}
g(r_{ij})=\frac{e^{ik_0r_{ij}}}{k_0r_{ij}},
\label{equ:couplingscalar}
\end{equation}
where we used $\langle\frac{r_{ij,\alpha}r_{ij,\beta}}{r_{ij}^{2}}\rangle=\frac{1}{3}\delta_{\alpha\beta}$.

This result is also obtained in the scalar case, where the atoms are coupled to a scalar radiation field  \cite{PRL1,PRA1}.

\section{Collective emission rates}
\label{CER}

The reduced atomic density operator $\rho$, obtained from tracing the density operator
over the radiation degrees of freedom, obeys the following equation \cite{zoller}:
 \begin{equation}
\begin{split}
    \frac{\text{d}\rho}{\text{d}t}=-\frac{i}{\hbar} (H_{eff}\rho-\rho H_{eff}^\dag)\\+\Gamma_0\sum_{ij}\sum_{\alpha\beta}\Lambda_{\alpha\beta}(\textbf{r}_{ij}) S^{(-)}_{j,\alpha}{\rho}S^{(+)}_{i,\beta},
       \label{eq:master}
\end{split}
 \end{equation}
where the effective Hamiltonian (\ref{equ:effectiveHamiltonian}) includes both energy-conserving and virtual processes.
The random matrix \newline $\Lambda_{\alpha\beta}(\textbf{r}_{ij}) $ is the imaginary part of the resonant dipole-dipole interaction (\ref{equ:couplingvect}), namely, $\Lambda_{\alpha\beta}(\textbf{r}_{ij})\equiv\mbox{Im}[g_{\alpha\beta}(\textbf{r}_{ij})]$.

The cooperative spontaneous emission rates of "real" (detectable) photons from the gas are determined  by the time evolution of the ground state population obtained from  the reduced atomic density matrix \cite{ernst,tallet1,tallet2}:
\begin{equation}
   \frac{\text{d}\rho_{GG}}{\text{d}t}=\Gamma_0\sum_{ij}\sum_{\alpha\beta}\Lambda_{\alpha\beta}(\textbf{r}_{ij})\langle G| S^{(-)}_{j,\alpha}{\rho} S^{(+)}_{i,\beta}| G\rangle,
       \label{eq:GroundstateEvol}
\end{equation}
where the ground state population is $\rho_{GG}=\langle G|\rho|G\rangle$ and $|G\rangle=|g_1,...,g_i,...,g_N\rangle$.
As expected, these emission rates depend only on the imaginary part of the resonant dipole-dipole interaction which represents energy-conserving processes.

%The eigenvalue equation of $\Lambda_{\alpha\beta}(\textbf{r}_{ij})$ is
%\begin{equation}
% \sum_{j=1}^{N}\sum_{\beta}\Lambda_{\alpha\beta}(\textbf{r}_{ij})u_{j,\beta}^{(n)}=\Gamma_{n}u_{i,\alpha}^{(n)},
%       \label{eq:PER}
%\end{equation}
We denote the $n$-th dimensionless eigenvalue of  $\Lambda_{\alpha\beta}(\textbf{r}_{ij})$ by $\Gamma_{n}$  and the associated $n$-th eigenfunction by $u_{i,\alpha}^{(n)}$.
Using the orthonormality of the eigenfunctions, we can rewrite (\ref{eq:GroundstateEvol}) as
\begin{equation}
 \frac{\text{d}\rho_{GG}}{\text{d}t}= \Gamma_{0}\sum_{n=1}^{3N}\Gamma_{n}\langle G|S_n^{(-)} \rho S_n^{(+)}|G\rangle\label{eqc25},
\end{equation}
 where the  collective raising and lowering operators are
$S_{n}^{(\pm)}=\sum_{i=1}^{N}\sum_\alpha u_{i,\alpha}^{(n)}S_{i,\alpha}^{(\pm)}$.

Thus, it is possible to interpret the eigenvalues of the random matrix $\Lambda_{\alpha\beta}(\textbf{r}_{ij})$ as the cooperative spontaneous emission rates (in units of $\Gamma_0$) of "real" photons from the cloud , i.e., the photon escape rates from the atomic gas \cite{tallet1,tallet2}.

 In the vectorial case, the $3N \times 3N$ coupling matrix is
\begin{equation}
\begin{split}
\Lambda_{\alpha\beta}(\textbf{r}_{ij})=\frac{3}{2}\bigg[\left(\frac{\sin(k_0r_{ij})}{k_0r_{ij}}+\frac{\cos(k_0r_{ij})}{(k_0r_{ij})^2}-\frac{\sin(k_0r_{ij})}{(k_0r_{ij})^3}\right)\delta_{\alpha\beta}\\-\left(\frac{\sin(k_0r_{ij})}{k_0r_{ij}}+3\frac{\cos(k_0r_{ij})}{(k_0r_{ij})^2}-3\frac{\sin(k_0r_{ij})}{(k_0r_{ij})^3}\right)\frac{r_{ij,\alpha}r_{ij,\beta}}{r_{ij}^{2}}\bigg].
\end{split}
\label{eq:PERvect}
\end{equation}
This matrix is comprised of $N \times N$ blocks, where the dimension of each block is $3 \times 3$.

In the scalar case  the $N \times N$ coupling matrix is given by
\begin{equation}
\Lambda({r}_{ij})=\frac{\sin(k_0r_{ij})}{k_0r_{ij}}.  \label{eq:PERscal}
\end{equation}

The properties of (\ref{eq:PERscal}) have been studied in detail, both numerically \cite{PRL2} and analytically \cite{PRL2,skip3}. In the following sections we study numerically the vectorial coupling matrix (\ref{eq:PERvect}) and compare the obtained results to those of the scalar case.

The average density of eigenvalues of  $\Lambda$  is
\begin{equation}
 P(\Gamma)=\frac{1}{M}\overline{\sum_{n=1}^{M}\delta(\Gamma-\Gamma_{n})}\label{eqc30},
\end{equation}
where $M=3N$ (resp. $N$) in the vectorial (resp. scalar) case.
The average, denoted by $\overline{\cdot
\cdot \cdot}$, is taken over the spatial random configurations of the atoms.

We examine two limiting cases.
In the dilute gas limit ($k_{0}r_{ij}\gg1)$, $\Lambda_{\alpha\beta}(\textbf{r}_{ij})=\delta_{ij}\delta_{\alpha\beta}$ and $\Lambda({r}_{ij})=\delta_{ij}$. Thus,

\begin{equation}
P(\Gamma)=\delta(\Gamma-1), \label{DGL}
\end{equation}
and the single-atom spontaneous emission rate, $\Gamma_0$, is recovered in both the vectorial and scalar cases.

In the so-called Dicke limit ($k_{0}r_{ij}\ll1$),  $\Lambda_{\alpha\beta}(\textbf{r}_{ij})=\delta_{\alpha\beta}$ and $\Lambda({r}_{ij})=1$. Therefore, for both the vectorial and scalar cases,

\begin{equation}
P(\Gamma)=\frac{1}{N}[\delta(\Gamma-N)+(N-1)\delta(\Gamma)].
\label{PDL}
\end{equation}
Here, $\Gamma = N$ is the non-degenerate superradiant mode, while $\Gamma =0$ is the $(N-1)$-degenerate subradiant mode.

It should be noted that  the mean value of $\Gamma$ hardly  characterizes $P(\Gamma)$  since $\Gamma_{mean}=[\mbox{Tr}(\Lambda)]/M$ and $\mbox{Tr}(\Lambda)=M$ in both the vectorial and scalar cases, thus  $\Gamma_{mean}=1$ regardless of the system parameters.

Therefore, in order to characterize $P(\Gamma)$  and obtain a measure of photon localization we use  the  following  function:

\begin{equation}
 C=1-2\int_1^\infty d\Gamma P(\Gamma), \label{C}
\end{equation}
normalized to unity. The function $C$  measures the relative number of states having a vanishing escape rate.
In the dilute gas limit, (\ref{DGL}) implies that $C=0$, indicating photon delocalization \cite{note}.
In the Dicke limit, (\ref{PDL}) gives
\begin{equation}
C=1-\frac{2}{N}. \label{CDL}
\end{equation}
Thus, for $N\gg1$, $C=1$ and photons are localized in the gas.

In the scalar case, away from the Dicke limit, the function $C$  has been thoroughly studied. It exhibits a scaling behavior over a broad range of system size and density in one \cite{EPL}, two \cite{PRA2}, and three dimensions  \cite{PRL2}. In three dimensions, $C$ can be approximated asymptotically by
\begin{equation}
C\simeq 1-3\frac{N_\perp}{N}, \label{CDLA}
\end{equation}
where $N_\perp\equiv(k_0L)^{2}/4$ is the number of transverse photon modes in an atomic volume $L^3$.

%Alternatively, in the scalar case, the scaling function can be written as
%\begin{equation}
%C(s)=\frac{s^2}{\gamma(s)^2},
%\label{CaWSingle}
%\end{equation}
%with
%\begin{equation}
%s=\frac{\gamma}{\tanh(\gamma)}-1,
%\label{Single}
%\end{equation}
%where  $\gamma$ is a function of $N/N_\perp$. This description is based on a mapping of the cooperative emission of $N$ randomly
%distributed atoms onto a stochastic Markov process
%on a one-dimensional lattice with $N$ sites \cite{PRL2}. In light of the link between the behavior of $C$ and "small world" networks, we follow \cite{ScalingParam2}  and assume the parameter $\gamma$  scales as
%$\gamma=\sqrt{4a\frac{N}{N_{\perp}}+(a\frac{N}{N_{\perp}})^2}$, where $a\simeq0.60$ is a numerical constant.

In the next sections we  study $C$ in the vectorial case and show that it exhibits a scaling behavior and obeys (\ref {CDLA}) as well.

\section{Method}
\label{Met}

In order to  obtain  $P(\Gamma)$ and $C$ in the vectorial case beyond the two limits discussed in the previous section, we follow  \cite{PRL2}. We consider $N \gg 1$ atoms enclosed in a cubic volume $L^{3}$. The atoms are distributed with a uniform density $\rho=N/{L}^{3}$. Using the resonant radiation wavelength, $\lambda=2\pi/ k_{0}$, we define the dimensionless density $\rho \lambda^3$.
Next, we introduce the Ioffe-Regel number \cite{IR}, $k_0l$, where $l$ is the photon elastic mean free path, namely $l=1/\rho\sigma$ and $\sigma$ is the average single scattering cross section. For resonant scattering, the  scattering cross section varies as $\lambda^2$, so that the Ioffe-Regel number can be written as  $k_0l^{(s)}=2\pi^2/ \rho \lambda^3$ in the scalar case and $k_0l^{(v)}=(2 / 3)  k_0l^{(s)}$ in the vectorial case \cite{book}.
Finally, we define the (on resonance) optical thickness, $b_0$, as the ratio between the system size $L$ and the photon elastic mean free path $l$. Using the definitions above, one obtains $b_0^{(s)}=N^{1/3} (\rho \lambda^3)^{2/3} / \pi$ and
$b_0^{(v)} = (3/2) b_0^{(s)}$. It is important to note that the optical thickness is related to the number of transverse photon modes by $b_0^{(s)}=\pi N/N_\perp$ and $b_0^{(v)} = (3 \pi/2) N/N_\perp$. Thus, $b_0$ is the scaling parameter in  (\ref{CDLA}).

While the Ioffe-Regel number accounts for disorder effects \cite{IR2}, cooperative effects are better described by the optical thickness \cite{bienaime,PRL2,lin1,lin2}. Therefore, we will use these two parameters  to investigate the distinctive roles of disorder and  cooperative effects in atomic gases.

For a given spatial atomic configuration, we numerically calculate the $3N$ eigenvalues of (\ref {eq:PERvect}) in the vectorial case, and, for comparison, the $N$ eigenvalues of (\ref {eq:PERscal}) in the scalar case. By varying the spatial configuration of the scatterers and averaging over disorder we obtain (\ref {eqc30}) and (\ref {C}).

\section{Photon escape rate distribution}
\label{Res1}

In this section we study  $P(\Gamma)$ in the vectorial case and compare the results to the scalar case, both for the large sample regime $(L>\lambda)$ and the small sample regime $(L\ll\lambda)$.
% The study of these distributions allows to obtain important information on the transport properties of the system \cite{KottosPRL,KottosPRB,orlowski}.

\subsection{Large sample regime}
\label{PERD}

\begin{figure}[h!]
\centering
\begin{minipage}[b]{8.5cm}
\includegraphics[width=8.5cm]{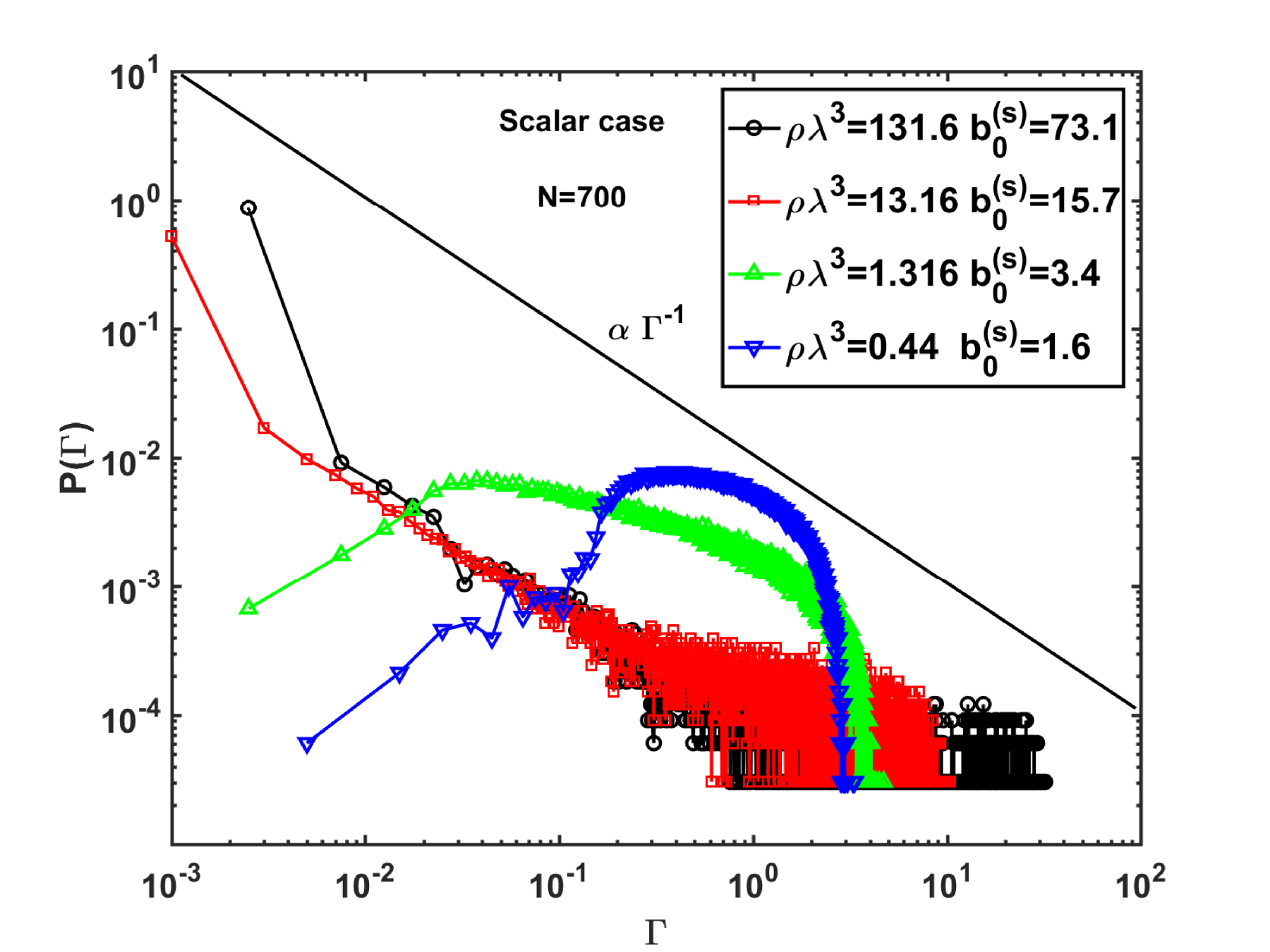}
\end{minipage}
\hspace{0.5cm}
\begin{minipage}[b]{8.5cm}
\includegraphics[width=8.5cm]{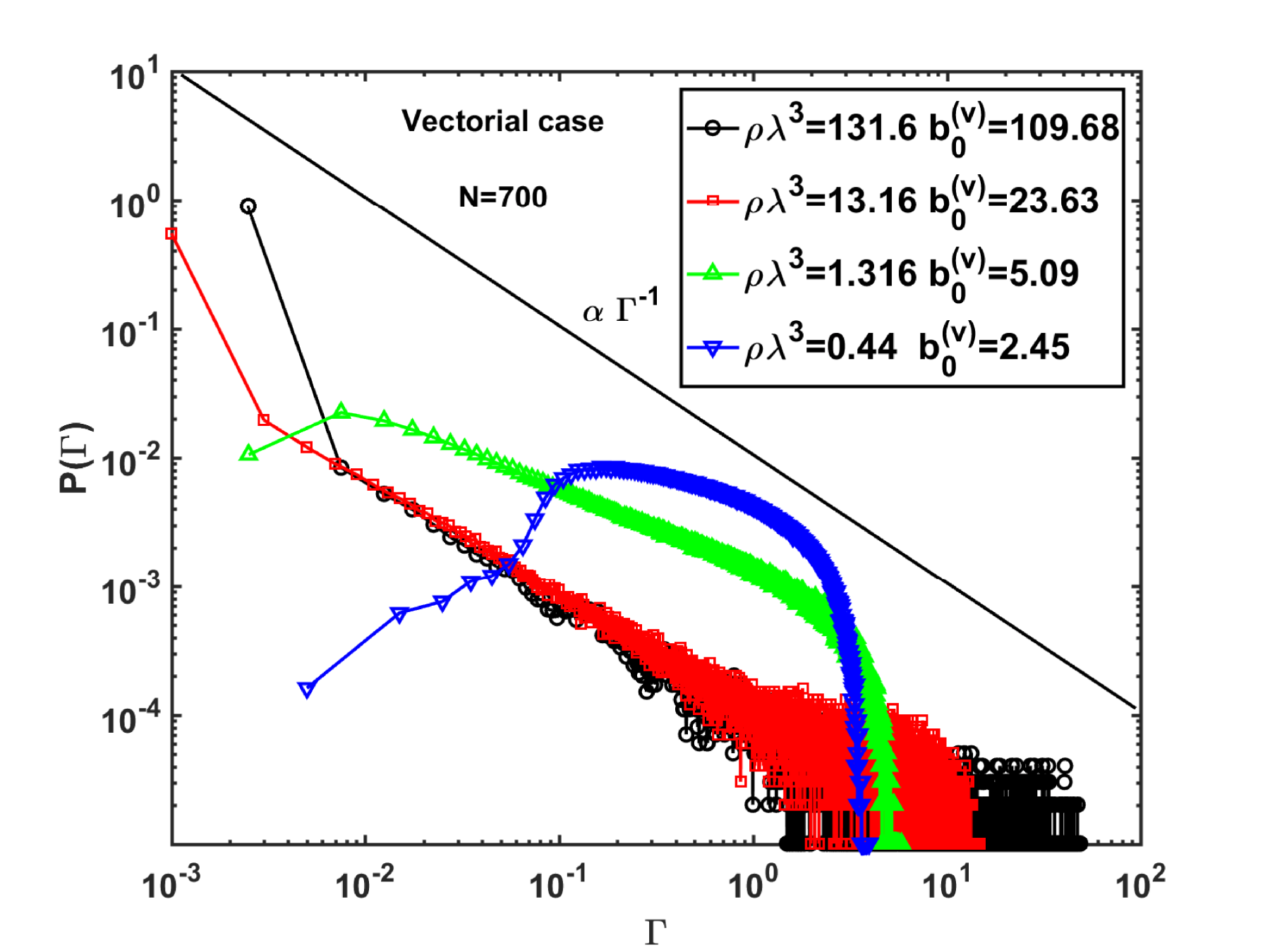}
\caption{\em (Color
   online) Photon escape rate distribution $P(\Gamma)$ (large sample regime) for $N=700$ atoms  in the scalar (top) and vectorial (bottom) cases for various cloud densities $\rho\lambda^3$.}
 \label{fig1}
\end{minipage}
\end{figure}

We first consider the photon escape rate distributions in the large sample size limit ($L>\lambda$). Figure \ref{fig1} shows the photon escape rate distribution $P(\Gamma)$ for a gas of $N=700$ atoms with increasing spatial densities. The results show that  the distribution is qualitatively the same for both the scalar and vectorial cases. For dilute gases, the distributions are peaked around the single atom decay rate, $\Gamma=1$, as predicted by (\ref{DGL}).

When increasing the density, we observe that $P(\Gamma)$ is shifted toward lower values of $\Gamma$, indicating the existence of long-living modes of the photon inside the sample. For dense clouds, when the  optical thickness is large enough, the distribution in both cases is  well described by the $P(\Gamma) \sim \Gamma^{-1}$ power law, as suggested for the scalar case in \cite{orlowski}.

\begin{figure}[h!]
\centering
\begin{minipage}[b]{8.5cm}
\includegraphics[width=8.5cm]{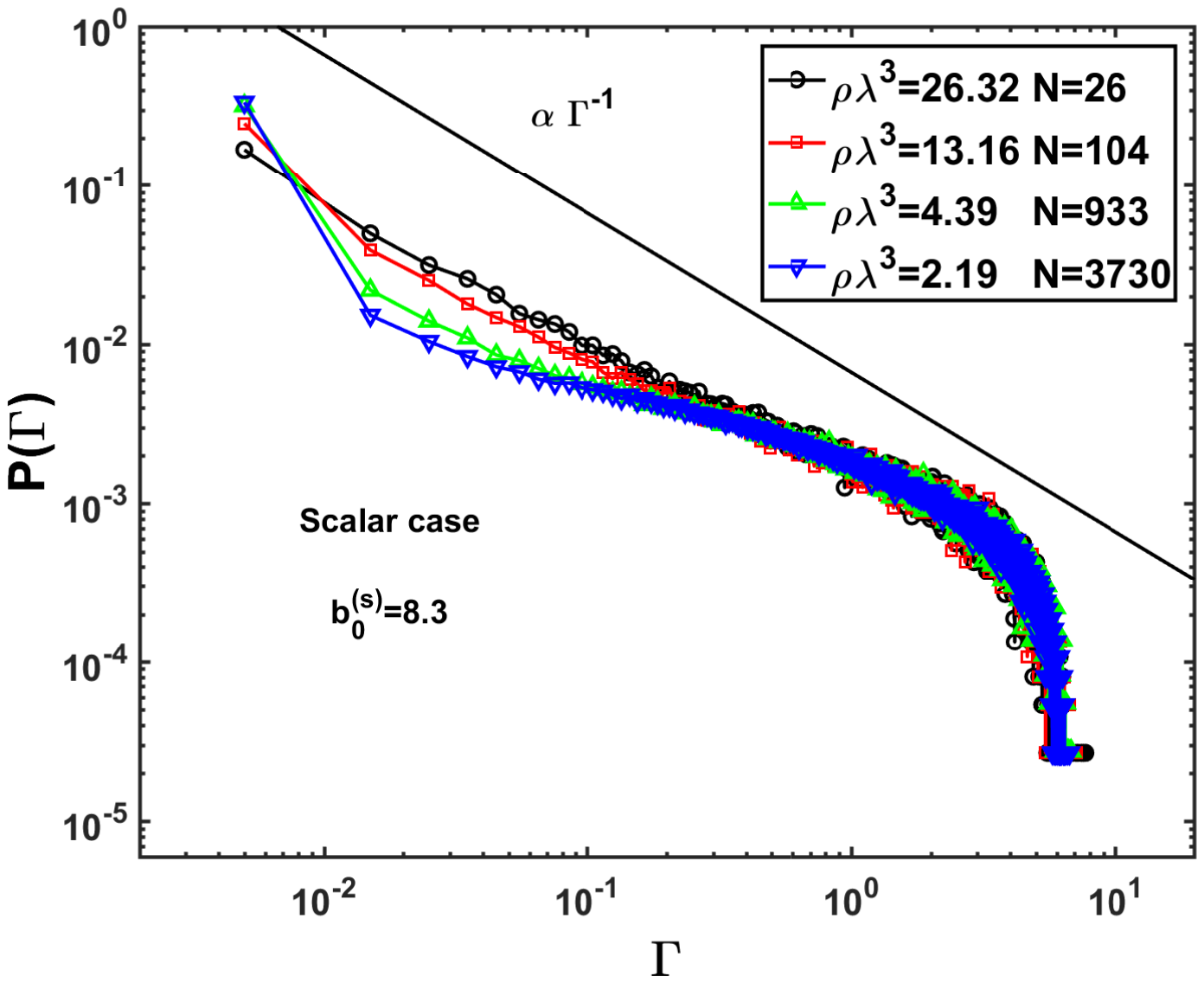}
\end{minipage}
\hspace{0.5cm}
\begin{minipage}[b]{8.5cm}
\includegraphics[width=8.5cm]{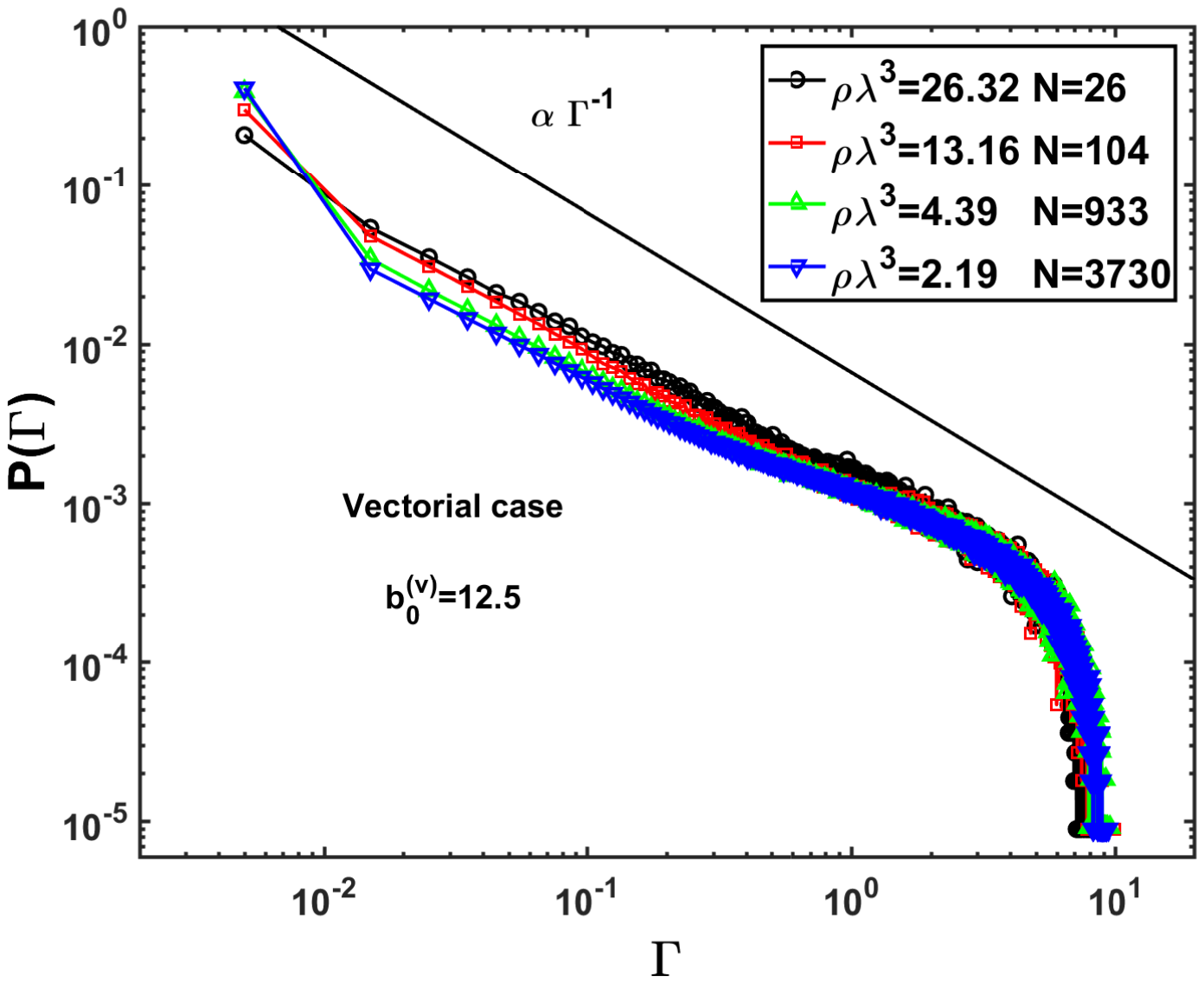}
\caption{\em (Color
   online) Photon escape rate distribution $P(\Gamma)$ (large sample regime) at a fixed optical thickness ($b_0^{(s)}=8.30$ and $b_0^{(v)}=12.50$) in the scalar (top) and vectorial (bottom) cases for various cloud densities $\rho\lambda^3$.}
 \label{fig2}
\end{minipage}
\end{figure}

Figure \ref{fig2} shows the distribution $P(\Gamma)$ at a fixed optical thickness for various spatial densities of scatterers. In both cases, for large enough densities, the photon escape rate obeys the power law of $P(\Gamma) \sim \Gamma^{-1}$.

\begin{figure}[h!]
\centering
\begin{minipage}[b]{8.5cm}
\includegraphics[width=8.5cm]{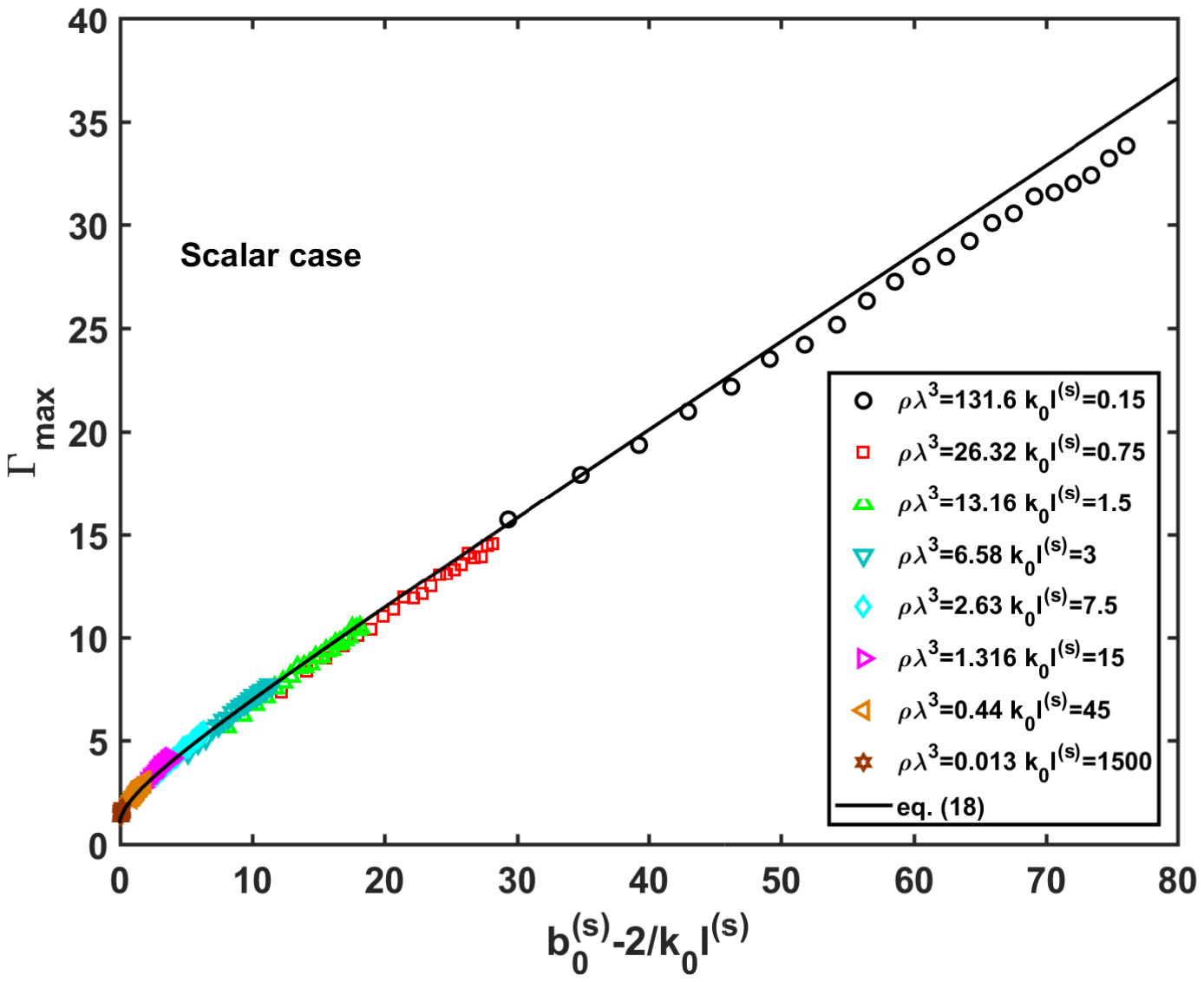}
\end{minipage}
\hspace{0.5cm}
\begin{minipage}[b]{8.5cm}
\includegraphics[width=8.5cm]{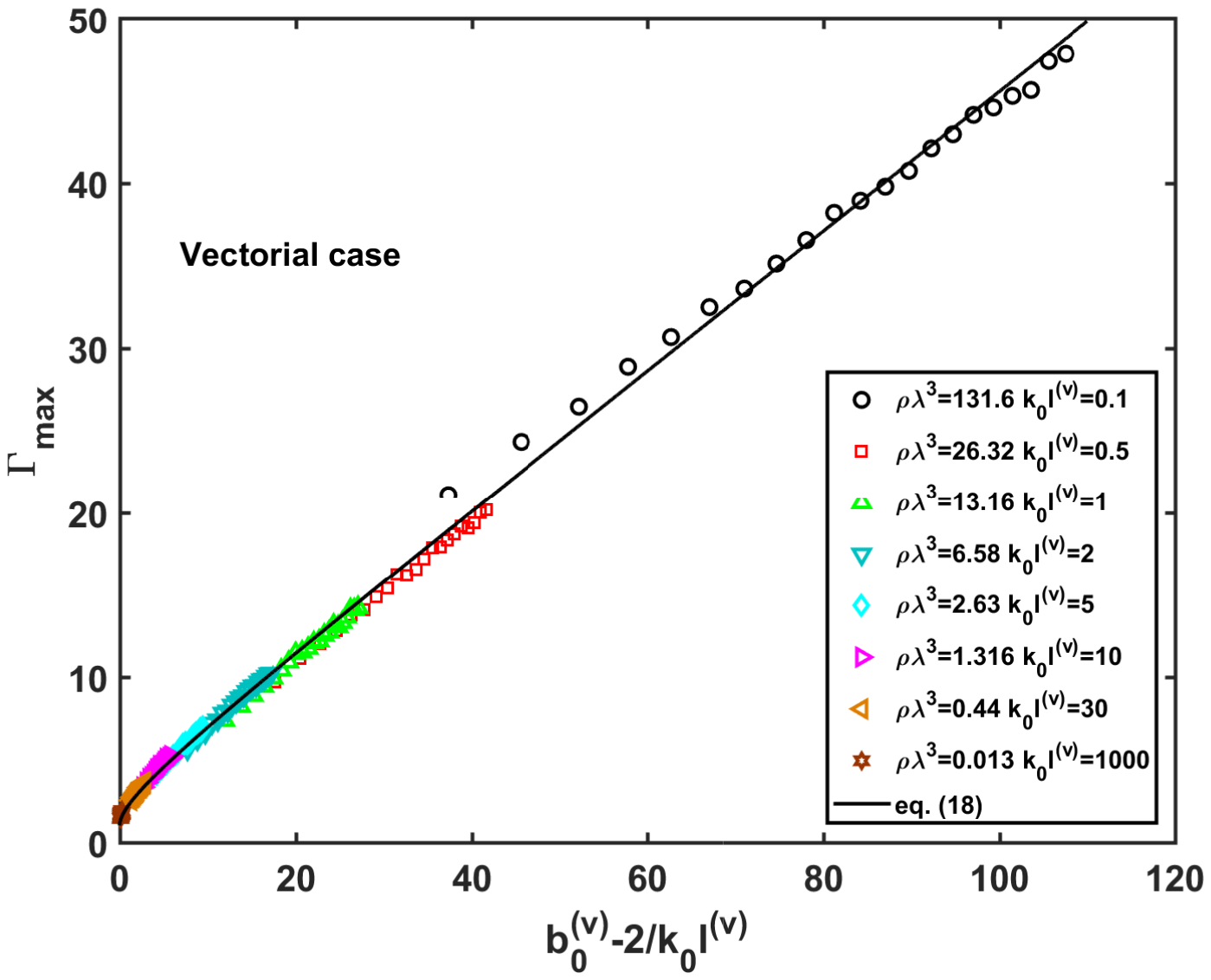}
\caption{\em (Color
   online) Maximal photon emission rate $\Gamma_{max}$ (large sample regime) in the scalar (top) and vectorial (bottom) cases. The solid line is given by (\ref{maxs}) in both
cases.}
 \label{fig3}
\end{minipage}
\end{figure}

Next, we consider the configuration-averaged maximal photon emission rate, $\Gamma_{max}$.
Figure \ref{fig3} shows the behavior of $\Gamma_{max}$  as a function of the optical thickness rescaled by the spatial density of the gas for both the scalar and vectorial cases. No qualitative differences between the scalar and vectorial cases are observed, and  in both cases the configuration-averaged maximal photon emission rate follows the expression:
\begin{equation}
\begin{split}
\Gamma^{(s,v)}_{max}=\sqrt{\frac{b_0^{(s,v)}-2/k_0l^{(s,v)}}{A}+\bigg(\frac{b_0^{(s,v)}-2/k_0l^{(s,v)}}{B}\bigg)^2}\\+\frac{b_0^{(s,v)}-2/k_0l^{(s,v)}}{C}+1,
\label{maxs}
\end{split}
\end{equation}
where $A=1.00$, $B=5.00$, and $C=4.50$ are free fitting parameters obtained numerically.
Equation (\ref{maxs}) results from the eigenvalue distribution of certain random matrices \cite{skip3} with some numerical adjustments \cite{BGAK}. It
indicates that $\Gamma_{max}$ is dominated by cooperative
effects, depending on the optical thickness, and slightly
corrected by disorder effects, depending on the spatial
density of the cloud. This equation recovers the asymptotic behavior predicted by the Marchenko-Pastur law \cite{skip3,MPL}, namely $\Gamma_{max}\propto \sqrt{b_0-2/k_0l}$ for dilute gases, and $\Gamma_{max}\propto b_0-2/k_0l$ for dense clouds.

\begin{figure}[h!]
\centering
\begin{minipage}[b]{9.0cm}
\includegraphics[width=9.0cm]{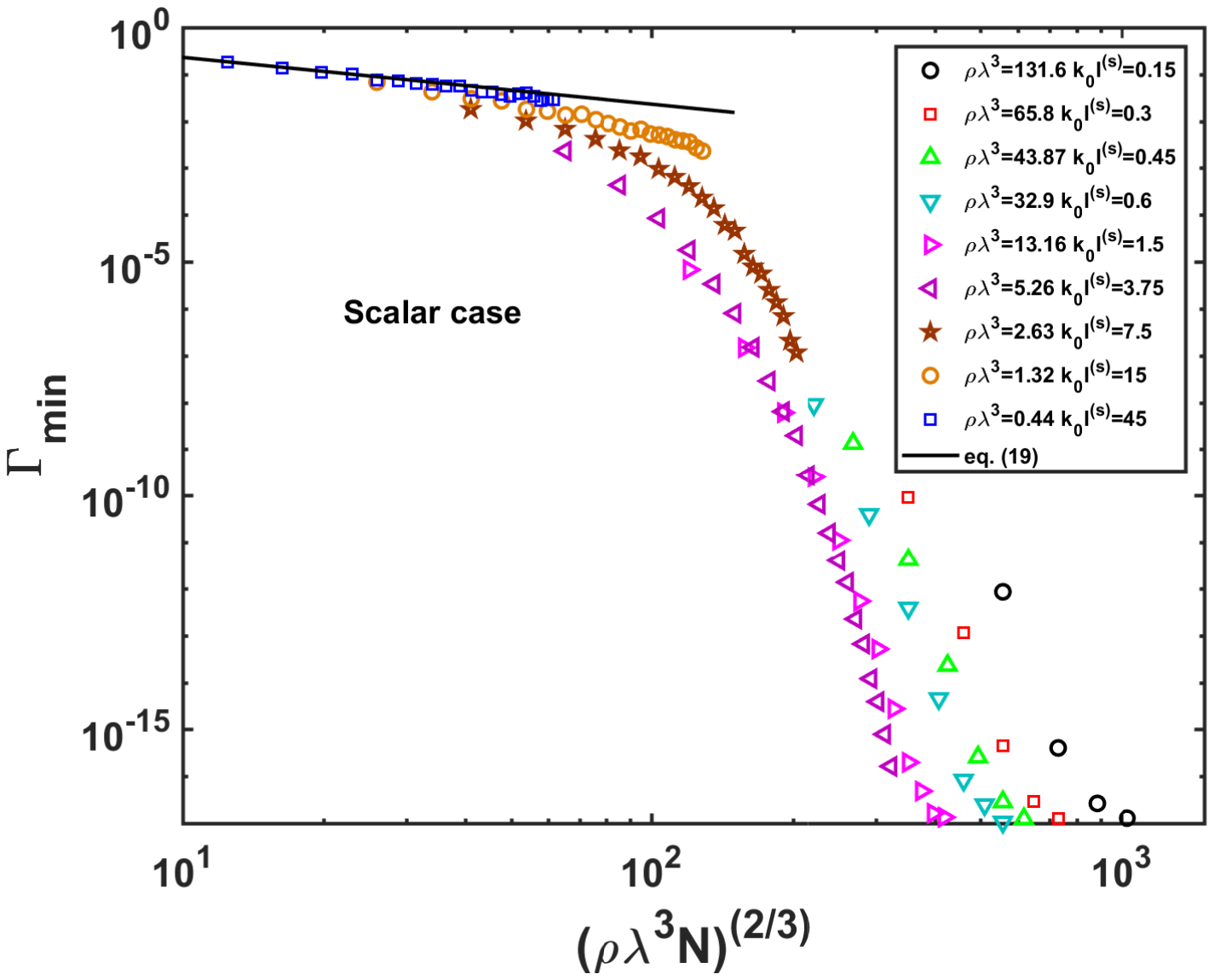}
\end{minipage}
\hspace{0.5cm}
\begin{minipage}[b]{9.0cm}
\includegraphics[width=9.0cm]{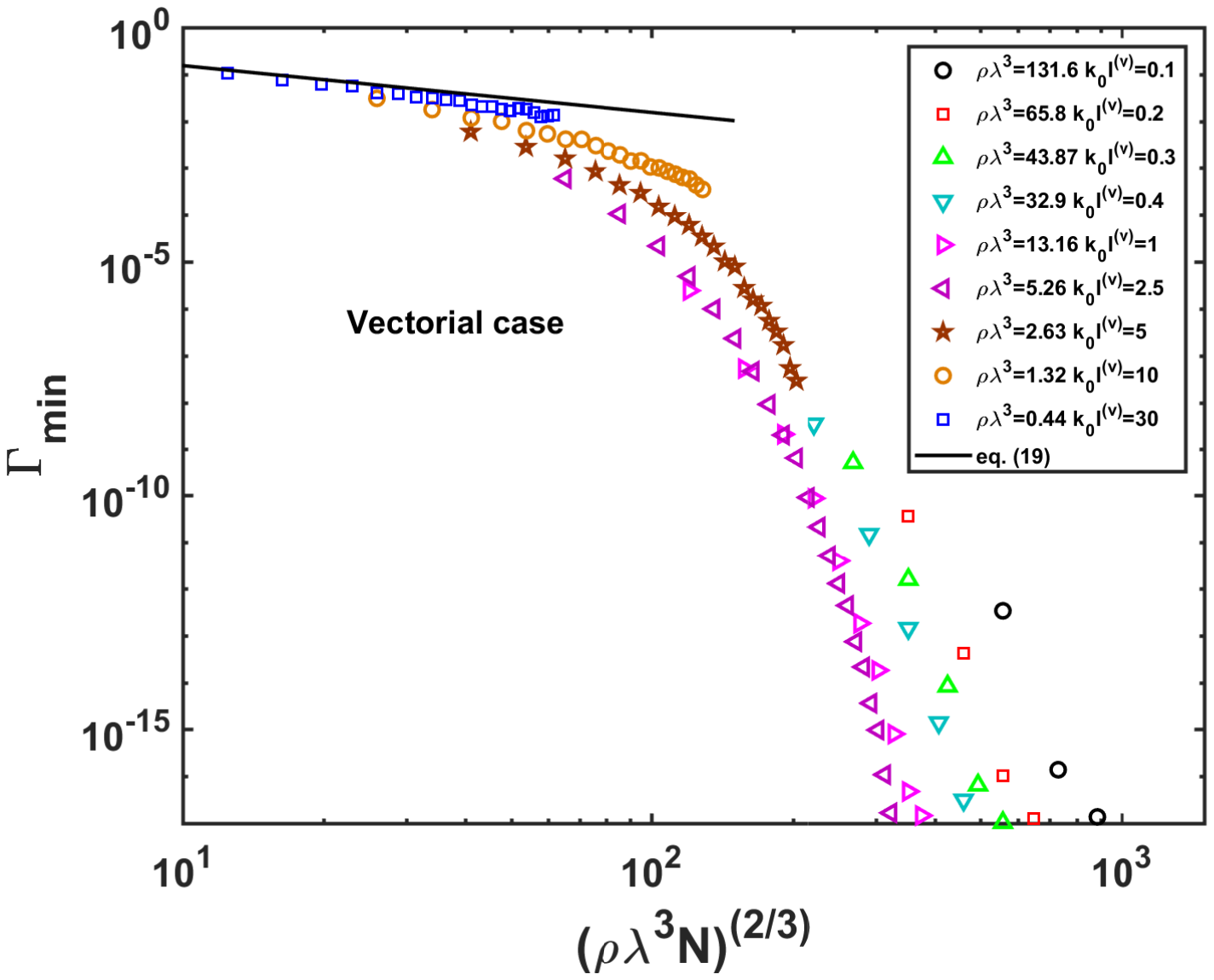}
\caption{\em (Color
   online) Minimal photon emission rate $\Gamma_{min}$ (large sample regime)  in the scalar (top)  and  vectorial (bottom) cases. The solid line in both cases  is given by (\ref{min}).}
 \label{fig4}
\end{minipage}
\end{figure}

Finally, we investigate the configuration-averaged minimal photon emission rate, $\Gamma_{min}$. Previous studies have shown that in dilute gases $\Gamma_{min}$ is dominated by subradiant pairs and is given by \cite{BGAK,skip3}:
\begin{equation}
\Gamma^{(s,v)}_{min}\propto (\rho\lambda^{3}N)^{-2/3}.
\label{min}
\end{equation}
As displayed in Fig. \ref{fig4}, both for the scalar and vectorial cases, equation (\ref{min}) holds for low densities but breaks down for dense gases.

\subsection{Small sample regime}

\begin{figure}[h!]
\centering
\includegraphics[width=8.5cm]{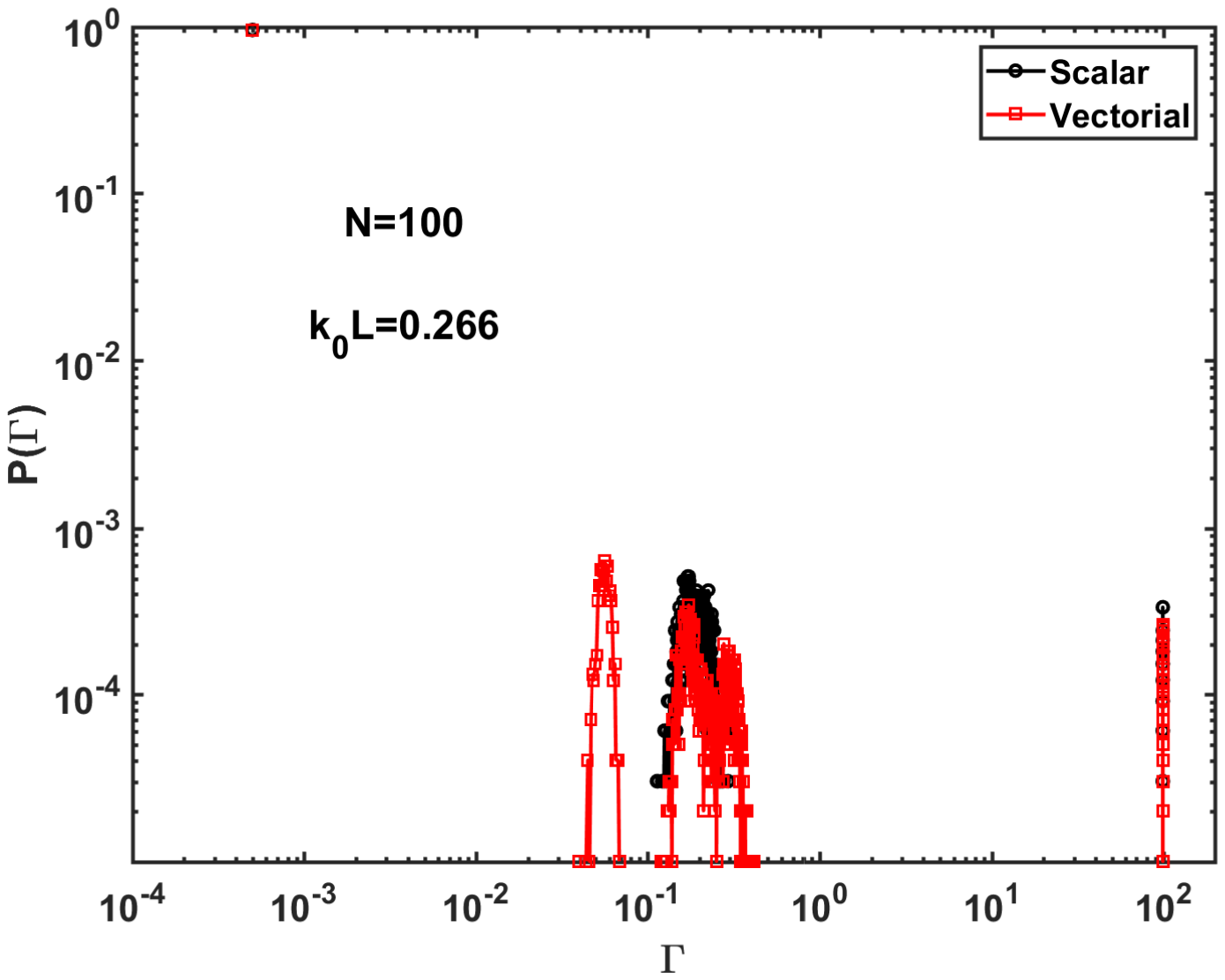}
\caption{\em (Color
   online) Photon escape rate distribution $P(\Gamma)$ in the scalar (black circles) and vectorial (red squares) cases for $N=100$ atoms enclosed in a system of size $k_0L=0.266$.}
 \label{fig5}
\end{figure}

Figure \ref{fig5} shows the photon escape rate distribution in the small sample regime for a cloud of  $N=100$ atoms enclosed in a system of size  $k_0L=0.266$ in the scalar and vectorial cases. In both cases, we observe the superradiant mode situated at $\Gamma=N$ and the subradiant modes close to $\Gamma=0$, as predicted by (\ref{PDL}).

\section{Scaling function}
\label{Res2}

\begin{figure}[h!]
\centering
\begin{minipage}[b]{8.5cm}
\includegraphics[width=8.5cm]{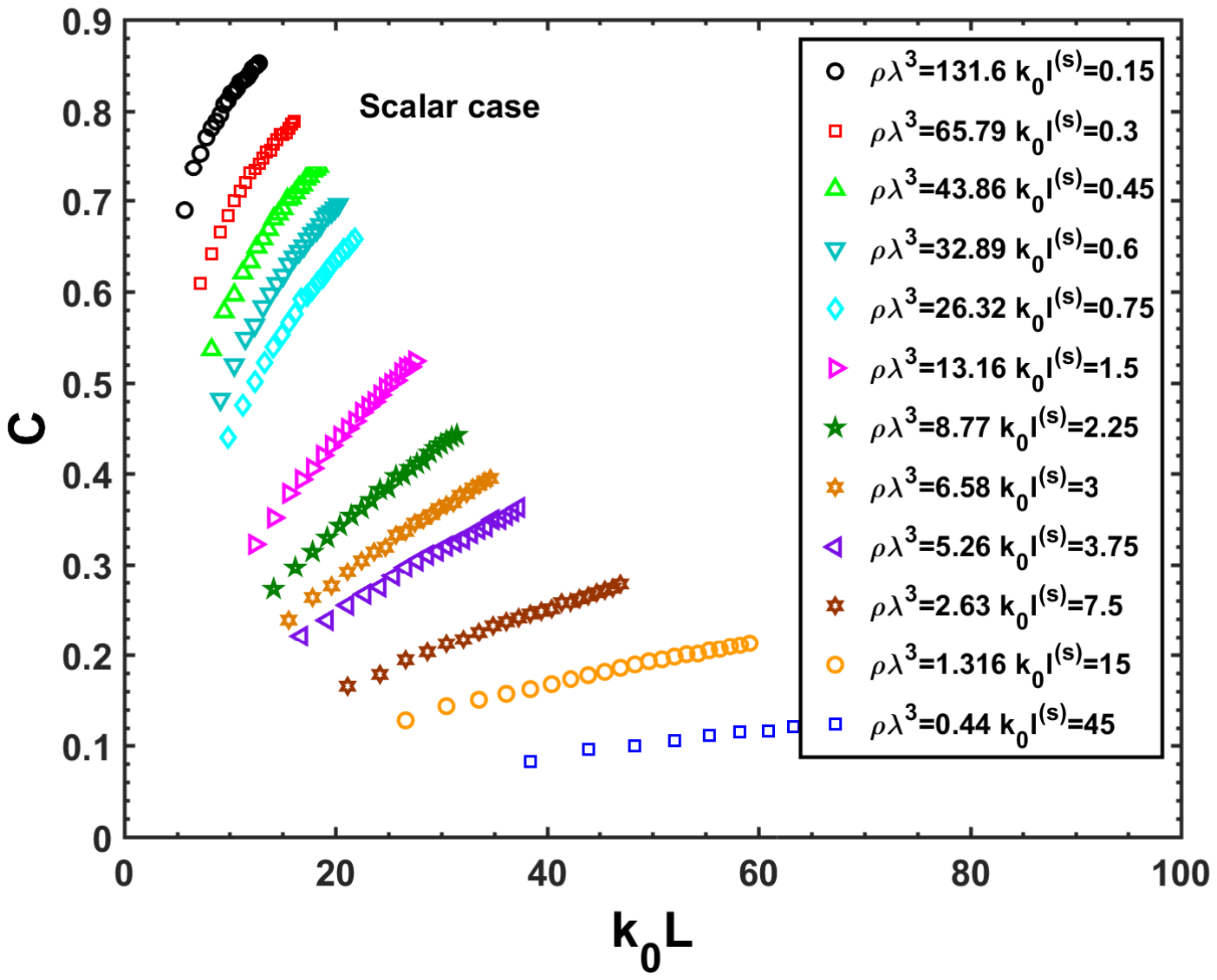}
\end{minipage}
\hspace{0.5cm}
\begin{minipage}[b]{8.5cm}
\includegraphics[width=8.5cm]{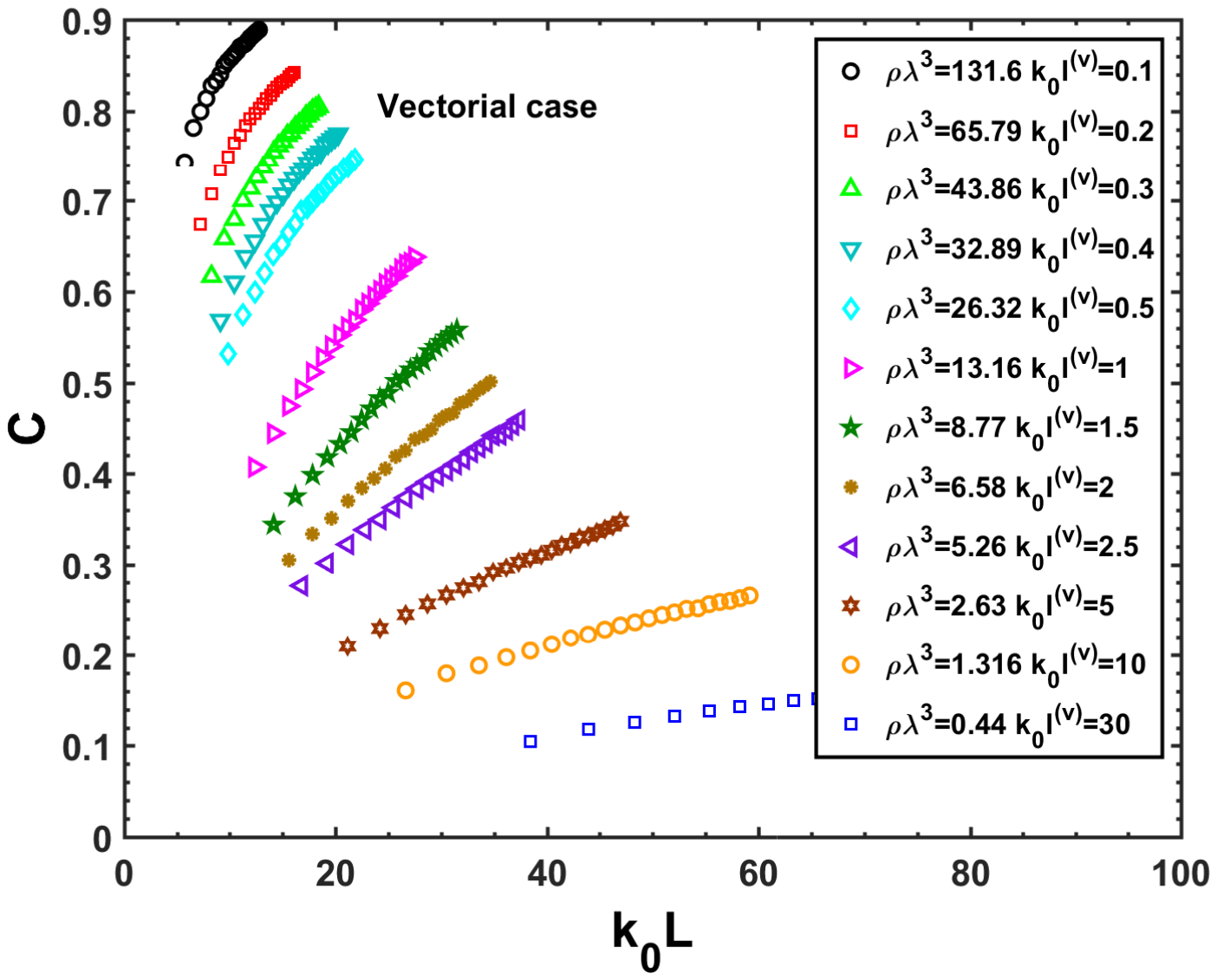}
\caption{\em (Color
   online) $C$ as a function of the system size $k_0L$ (large sample regime) for increasing atomic densities $\rho\lambda^3$  in the scalar (top) and vectorial (bottom) cases.}
 \label{fig6}
\end{minipage}
\end{figure}

Figure \ref{fig6} shows the behavior of  $C$ defined in (\ref{C}) as a function of the system size  $(L>\lambda)$ for increasing atomic densities in the scalar and vectorial cases. No significant differences are observed between the two cases.
For dilute (or optically thin) gases, $P(\Gamma)$ is peaked around the single atom decay rate, so that $C\rightarrow 0$. For dense (or optically thick) media, according to Section \ref{Res1},  $P(\Gamma)$ is  shifted toward lower values of $\Gamma$, hence $C\rightarrow 1$.

\begin{figure}[h!]
\centering
\includegraphics[width=8.5cm]{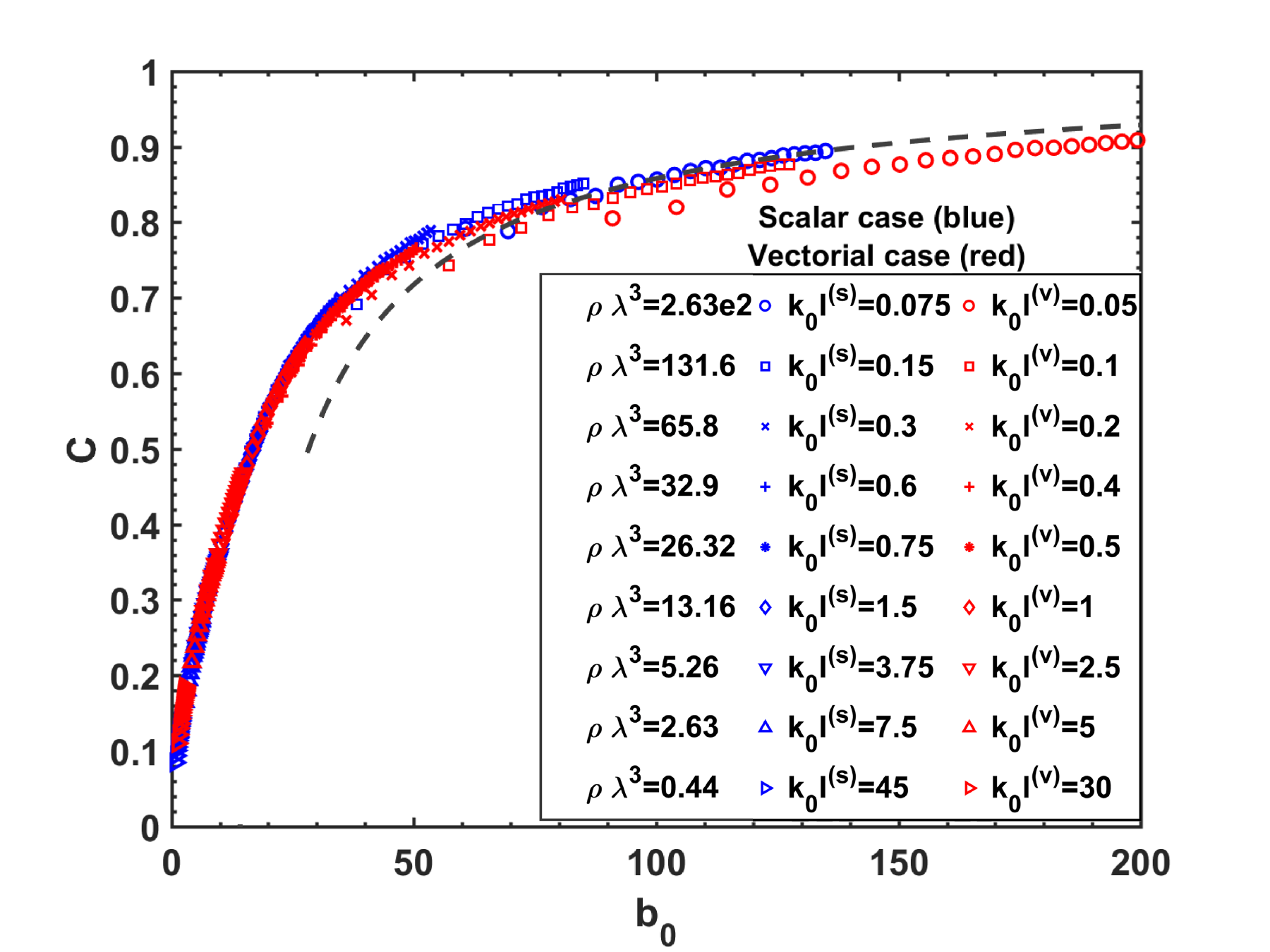}
\caption{\em (Color
   online) $C$ as a function of the optical thickness $b_0$ (large sample regime) for increasing atomic densities $\rho\lambda^3$  in the scalar (blue) and vectorial (red) cases. In both cases, the dashed line is given by (\ref{CDLA}).}
 \label{fig7}
\end{figure}

The data of  Fig. \ref{fig6} (both in the scalar and vectorial cases) collapse on a single curve (Fig. \ref{fig7})  when plotted as a function of the optical thickness, indicating that photons undergo a crossover from delocalization toward localization as the scaling variable $b_0$  is increased. Therefore, photon localization is dominated by cooperative effects rather than disorder. In both cases, the numerical data are in line with the theoretical expression (\ref{CDLA}).

The obtained result that the scaling function increases with the optical thickness can be explained as follows. As the number of scattering events is increased, it takes more time for the photon to leave the gas, thus the relative number of states having a vanishing escape rate is increased.

\begin{figure}[h!]
\centering
\includegraphics[width=8.5cm]{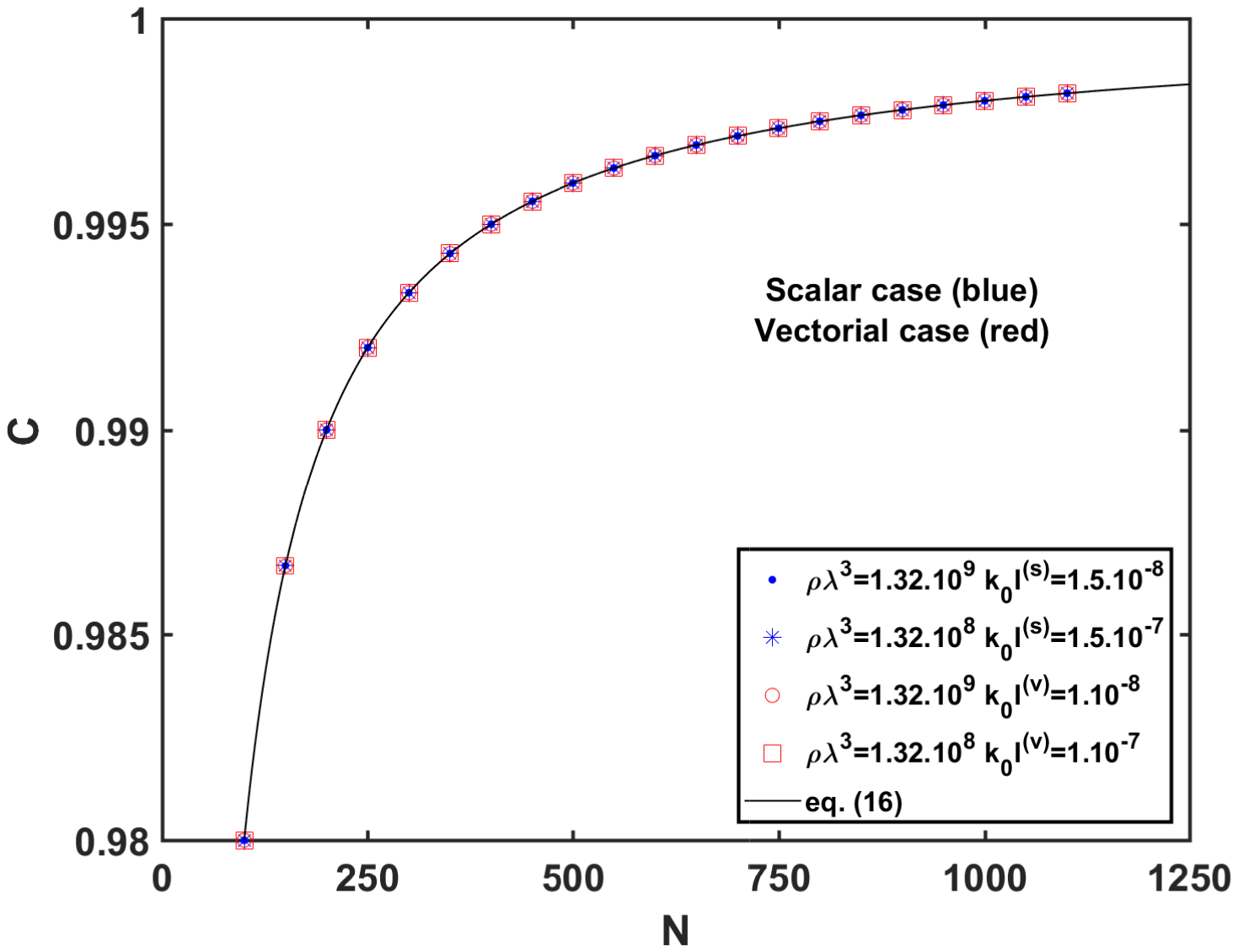}
\caption{\em (Color
   online) $C$ as a function of the number of atoms $N$ (small sample regime) for the scalar (blue) and vectorial (red) cases. The solid lines is given by (\ref{CDL}) in both cases. For a density of $\rho\lambda^3=1.32\cdot 10^{8}$, the system size is $0.06\le k_0L\le0.13$, and for a density of $\rho\lambda^3=1.32\cdot10^{9}$, the system size is $0.03\le k_0L\le0.06$.}
 \label{fig8}
\end{figure}

In the small sample regime $(L\ll\lambda)$, the behavior of the function $C$ is shown in Fig. \ref{fig8} for the scalar and vectorial cases. In both cases, the numerical data are in full agreement with the theoretical prediction (\ref{CDL}).

\begin{figure}[h!]
\centering
\includegraphics[width=9.0cm]{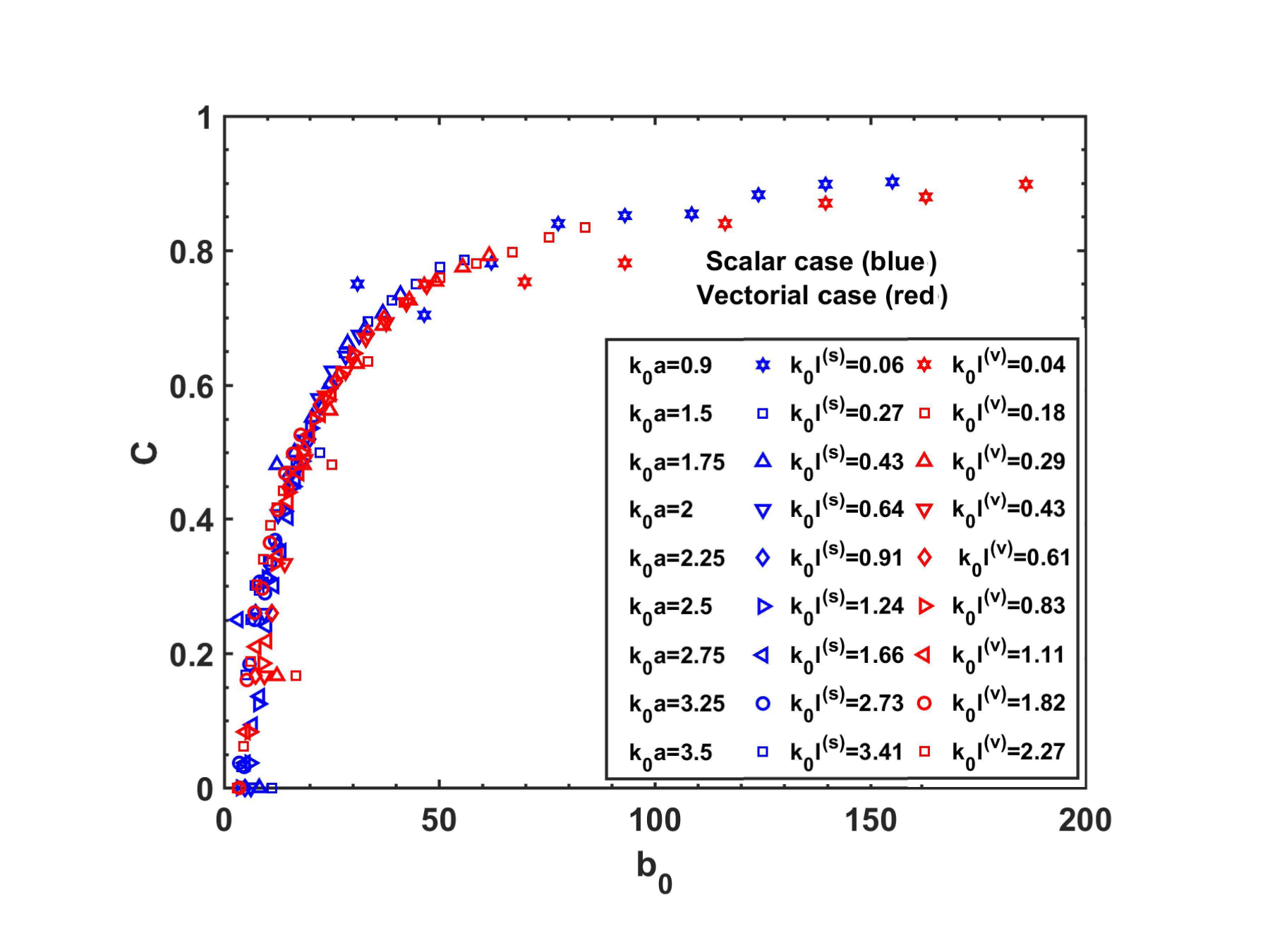}
\caption{\em (Color
   online) $C$ as a function of the optical thickness $b_0$ (large sample regime) in the case of an ordered medium, a cubic lattice of $8\le N\le1000$ atoms and an inter-atomic distance $a$. The scalar model is in blue and the vector model in red.}
 \label{fig9}
\end{figure}

In order to verify that the scaling behavior observed in Fig. \ref{fig7} is indeed dominated by cooperative effects rather than disorder, we study the quantity $C$ in the case of a spatially ordered medium (large sample regime). We consider a cubic lattice with an inter-atomic distance $a$. In this case, the Ioffe-Regel number is given by $k_0l^{(s)}=2\pi^2a^3/\lambda^3$  \cite{IR} and the optical thickness is $b_0^{(s)}=N^{1/3}\lambda^2/2\pi a^2$. Figure \ref{fig9} shows $C$ as a function of the optical thickness.  A similar scaling behavior is observed, both in the scalar and vector models, corroborating that photon localization is dominated by cooperative effects rather than disorder.

\section{Discussion}
\label{discuss}

We have studied the photon escape rate distribution from an ensemble of random atomic scatterers coupled to a scalar or vector radiation field. In our analysis, virtual processes are allowed, but only the on-shell photons that leave the cloud are taken into account for detection purposes.  We have shown that in both the scalar and vectorial cases, the results are qualitatively the same, namely  the photon escape rate distribution follows the $P(\Gamma) \sim \Gamma^{-1}$ power law, the maximal photon emission rate recovers the asymptotic behavior predicted by the Marchenko-Pastur law, and in dilute gases the minimal photon emission rate is dominated by subradiant pairs. Moreover, in these two cases, the  function $C$ exhibits a scaling behavior over a broad range of system parameters, where the optical thickness serves as the scaling parameter (in the large sample regime). This observation is in line with the findings reported by the authors of \cite{PRL2},  who originally calculated  the scaling  function in the scalar case. The finding that photons undergo a crossover from delocalization toward localization rather than a disorder-driven phase transition as expected from Anderson localization \cite{gang}, suggests that photon localization is dominated by cooperative effects rather than disorder. This conclusion is further validated by the finding that a similar scaling behavior is obtained in ordered media. It should be noted, however, that the specific contribution of each mechanism to light localization, namely, cooperativity or disorder, cannot be singled out \cite
{BGAK}.

The results show that the scalar model is an excellent approximation when considering real photon escape rates from atomic gases. As the scalar model is much simpler compared to the vectorial one, the former could be used, without losing substantial information.
Since averaging the vectorial coupling matrix $\Lambda_{\alpha\beta}(\textbf{r}_{ij})$ (\ref{eq:PERvect}) over the random orientations of the atoms leads to the scalar coupling matrix $\Lambda(r_{ij})$ (\ref{eq:PERscal}), our results indicate that $P(\Gamma)$ is not sensitive to the detailed location of the atoms.

It is interesting to compare our findings to those obtained when considering the emission of both "real" and virtual photons, and calculating the spectrum of $g_{\alpha\beta}(\textbf{r}_{ij})$ (\ref{equ:couplingvect}) in the vectorial case and $g(r_{ij})$  (\ref{equ:couplingscalar}) in the scalar case. When taking into account the real part of $g$ as well,  in both the scalar and vectorial cases the photon escape rate distribution follows the $P(\Gamma) \sim \Gamma^{-4/3}$ power law  (a criterion for strong localization of scalar waves within the Anderson model \cite{KottosPRL,KottosPRB}), and the maximal photon emission rate recovers the asymptotic behavior predicted by the Marchenko-Pastur law (although the exact expression is different in each case) \cite{BGAK}. However, the inclusion of virtual emission processes also leads to significant differences between the scalar and vector models. The minimal photon emission rate in the scalar case is substantially different from that in the vector model.  Moreover, localization of light can be achieved in random three-dimensional atomic media only for a scalar radiation field; it cannot be achieved when the vectorial properties of the electromagnetic wave are taken into account  \cite{skip1,BGAK}.

A possible explanation for the considerable difference between the two cases is as follows: The real part of $g$ is responsible for van der Waals dephasing, a phenomenon that is sensitive to the detailed location of the scatterers \cite{gross}. Therefore, when dealing with the spectrum of the effective Hamiltonian, the orientation-averaged scalar model is essentially different from the vector model. In the study described in this paper, only the imaginary part of $g$ is relevant and van der Waals dephasing does not play a role. Hence,  the detailed location of the atoms is less important and an orientation-averaged calculation constitutes a very good approximation.

In summary, two approaches have been proposed to study light localization in atomic gases. The first approach explores the complex spectrum of the effective Hamiltonian describing the atomic system \cite{BGAK,orlowski,svidzinsky2,bienaime}. There, in the case of a scalar radiation field, light localization occurs as a phase transition. However, when the vectorial properties of the electromagnetic wave are taken into account, photons are always delocalized \cite{skip1,BGAK}. The second approach, discussed in this paper, investigates photon escape rates determined by the time evolution of the ground-state population \cite{ernst,tallet1,tallet2,PRL2}. According to this approach, both for scalar and vector fields, photons undergo a crossover from delocalization toward localization. As this result is valid in both ordered and disordered media, it can be concluded that the observable studied within the framework of the latter approach, i.e., photon escape rates, is very weakly affected by disorder.

The inclusion (or exclusion) of additional symmetries can significantly modify disorder-driven phase transitions. For example, according to scaling theory of Anderson, electronic states in two-dimensional media should be localized \cite{gang}.  However, in the presence of  spin-orbit coupling, a metal-insulator transition does occur in these systems  \cite{SOC1,SOC2}. Indeed,  the medium discussed here is three-dimensional. Yet, it could be that polarization introduces an additional symmetry that suppresses the transition.  Whereas this effect is reflected in the behavior of quantity used in the first approach, the observable studied in the second one is not affected since it is hardly sensitive to disorder.

\section{Conclusions}
\label{conclusions}
We have numerically studied photon escape rates from  three-dimensional atomic gases, while taking  into account the vectorial nature of light. We have shown that the vector and scalar models qualitatively follow the same scaling law. In both cases,  photons undergo a crossover from delocalization toward  localization as the optical thickness of the cloud is increased. Therefore, photon localization is dominated by cooperative effects rather than disorder. This theoretical conclusion calls for the identification of experimental signatures allowing to distinguish between the two mechanisms of photon localization.

\section*{Author contribution statement}
L.B. performed the numerical simulations. L.B. and A.G. analyzed the data. E.A. and R.K. supervised the project, and all authors contributed to the writing of the paper.


\begin{thebibliography}{99}
\bibitem{note} The $\delta$ function is considered as a limiting case of an even function peaked at zero. Thus, the integral of $\delta(x)$ from 0 to infinity equals 1/2.
\end{thebibliography}
\end{document}